\documentclass[conference]{IEEEtran}
\IEEEoverridecommandlockouts
\usepackage{cite}
\usepackage{amsmath,amssymb,amsfonts}
\usepackage{algorithmic}
\usepackage{graphicx}
\usepackage{textcomp}
\usepackage{xcolor}

\usepackage{xurl}
\usepackage{multirow}

\def\BibTeX{{\rm B\kern-.05em{\sc i\kern-.025em b}\kern-.08em
    T\kern-.1667em\lower.7ex\hbox{E}\kern-.125emX}}
\begin{document}

\title{On Crossover Distance for Optical Wireless Satellite Networks and Optical Fiber Terrestrial Networks\\
}

\author{\IEEEauthorblockN{Aizaz U. Chaudhry and Halim Yanikomeroglu}
\IEEEauthorblockA{\textit{Department of Systems and Computer Engineering} \\
\textit{Carleton University}\\
Ottawa, ON K1S 5B6, Canada \\
\{auhchaud, halim\}@sce.carleton.ca}
}

\maketitle

\begin{abstract}
Optical wireless satellite networks (OWSNs) can provide lower latency data communications compared to optical fiber terrestrial networks (OFTNs). The crossover function enables to calculate the crossover distance for an OWSN and an OFTN. If the distance between two points on Earth is greater than the crossover distance, then switching or crossing over from the OFTN to the OWSN results in lower latency for data communications between these points. In this work, we extend the previously proposed crossover function for a scenario such that intermediate satellites (or hops) are incorporated between ingress and egress satellites in the OWSN for a more realistic calculation of the crossover distance in this scenario. We consider different OWSNs with different satellite altitudes and different OFTNs with different optical fiber refractive indexes, and we study the effect of the number of hops on the crossover distance and length of a laser inter-satellite link (LISL). It is observed from the numerical results that the crossover distance increases with an increase in the number of hops, and this increase is higher at higher satellite altitudes in OWSNs and lower refractive indexes in OFTNs. Furthermore, an inverse relationship between the crossover distance and length of a LISL is observed. With an increase in the number of hops, the length of a LISL decreases as opposed to the crossover distance.
\end{abstract}

\vspace{0.2cm}


\section{Introduction}
\par Laser inter-satellite links (LISLs) between satellites in upcoming very low Earth orbit (VLEO) and low Earth orbit (LEO) satellite constellations will create optical wireless satellite networks (OWSNs), also known as free-space optical satellite networks, in space \cite{b1}, \cite{b2}. SpaceX \cite{b3} and Telesat \cite{b4} are planning to equip their VLEO/LEO satellites in their Starlink and Lightspeed satellite constellations with laser communication terminals \cite{b5,b6,b7} to establish LISLs between satellites in these constellations to realize OWSNs. For example, SpaceX has mentioned in their 2016 FCC filing that their Starlink constellation will employ laser (or optical) inter-satellite links for seamless network management and continuity of service, which will also help Starlink to meet spectrum sharing constraints with other constellations. Furthermore, OWSNs arising from such satellite constellations will help bridge the digital divide by delivering broadband Internet to unserved and underserved communities in rural and remote areas \cite{b8}. They will also be an ideal solution for low-latency communications for high-frequency trading between stock markets around the world, where a one millisecond advantage in latency can translate into \$100 million a year for a brokerage firm \cite{b9}. 

\par Data communications between satellites over LISLs in OWSNs takes place at the speed of light in space and the higher speed of light in the vacuum of space gives OWSNs an insurmountable advantage over optical fiber terrestrial networks (OFTNs) on Earth in terms of latency (or propagation delay). The speed of light in optical fiber in OFTNs is around 50\% lower than that in LISLs in OWSNs \cite{b10}. Also, data communications over long distances between source and destination ground stations has to bounce back and forth between satellites in space and intermediate ground stations on Earth without LISLs and this negatively affects the latency of an OWSN \cite{b11}, \cite{b12}.  

\par OWSNs can offer lower latency than OFTNs for long-distance inter-continental data communications between ground stations in different cities \cite{b13}. However, when should the data traffic switch or crossover from an OFTN on Earth to an OWSN in space for lower latency data communications between two points, such as optical fiber relay stations or satellite ground stations, on the surface of the Earth? Crossover functions have been proposed to calculate the crossover distance, i.e., the distance between two points on Earth beyond which switching or crossing over from an OFTN to an OWSN leads to lower latency data communications between these points, in different scenarios \cite{b14}. In this work, we extend one of these crossover functions. More specifically, the contribution of this work is as follows. \emph{We enhance the crossover function for the first scenario in \cite{b14} to incorporate intermediate satellites (or hops) between the ingress and egress satellites in the OWSN for a more realistic determination of the crossover distance in this scenario.}

\par We evaluate the effect of different number of hops on the crossover distance and length of a LISL, and we consider three different OWSNs with different satellite altitudes and three different OFTNs with different optical fiber refractive indexes for this study. Here, we define the number of hops as the number of intermediate satellites between the ingress and egress satellites in an OWSN. The crossover distance varies with the propagation distance between ingress and egress satellites in an OWSN, and the numerical results indicate that this propagation distance increases with an increase in the number of hops between ingress and egress. An increase in the number of hops means more LISLs, which result in a higher propagation distance between ingress and egress, and this translates into a higher crossover distance. 

\par We further observe that the increase in the crossover distance with an increase in the number of hops is higher at higher satellite altitudes and lower optical fiber refractive indexes since the increase in propagation distance between ingress and egress is higher at these satellite altitudes and refractive indexes. An inverse relationship between the crossover distance and length of a LISL is seen. As the number of hops increases for a certain satellite altitude and optical fiber refractive index, the length of a LISL decreases as there are more, indirect, and shorter LISLs between ingress and egress, which lead to an increase in the propagation distance between ingress and egress and, thereby, an increase in the crossover distance.

\par The rest of the paper is organized as follows. Section II briefly discusses the related work and the motivation for this work. The details of incorporating intermediate satellites in the crossover distance are given in Section III. Section IV presents the numerical results and their explanation. The paper concludes in Section V with some discussion on future work.

\section{Related Work}

\par The comparison of satellite networks arising from VLEO/LEO satellite constellations and optical fiber terrestrial networks in terms of latency has been investigated in the literature \cite{b1,b13,b14,b15,b16}. A hypothetical satellite constellation, consisting of 1,600 satellites at 550 km altitude, was considered in a study that observed an improvement of 70\% in the round-trip time over the satellite network as compared to Internet latency \cite{b15}. Another study used Starlink’s original Phase I satellite constellation of 1,600 satellites at 1,150 km altitude and showed that the corresponding satellite network could provide lower latency as compared to the OFTN for data communications over longer distances \cite{b16}.

\par The suitability of OWSNs for providing low-latency communications over long distances was investigated in comparison with OFTNs by considering an OWSN operating at 550 km altitude \cite{b1}. It was shown that the OWSN outperformed the OFTN in terms of latency for data communications over distances longer than 3,000 km. Starlink’s Phase I constellation at 550 km altitude was used to conduct a comparison of an OWSN and an OFTN in terms of latency in different scenarios for long-distance inter-continental data communications \cite{b13} and it was observed that the OWSN outperformed the OFTN in all scenarios.

\par In \cite{b14}, crossover functions were proposed to enable the calculation of the crossover distance in four different scenarios. It was concluded that the crossover distance depended upon the altitude of satellites in the OWSN, the optical fiber refractive index in the OFTN, and the end-to-end propagation distance over the OWSN, i.e., the propagation distance between source and destination points (or satellite ground stations) in different cities over the OWSN. However, a direct LISL was assumed to exist between the ingress and egress satellites, which is usually not the case in OWSNs arising from real satellite constellations. In this work, we extend the crossover function for the first scenario proposed in \cite{b14} to conform it to a more realistic scenario to enable a more realistic determination of the crossover distance by incorporating intermediate satellites between ingress and egress.

\section{Incorporating Intermediate Satellites in Crossover Distance}

\par The crossover functions to determine the crossover distance have been introduced for four different scenarios in \cite{b14}. These crossover functions assume a direct LISL between ingress and egress satellites in an OWSN, which is generally not the case in reality. In real scenarios, there could be one or more intermediate satellites between ingress and egress satellites in an OWSN. In this work, we incorporate intermediate satellites between ingress and egress in the crossover function for the first scenario in \cite{b14}, and subsequently study the effect of varying the number of intermediate satellites on the crossover distance and length of a LISL in this scenario. The crossover function for this scenario, which is shown in Fig. 1, is derived in \cite{b14} as

\begin{equation} \tag{1}
	\label{eq_1}
	f_{crossover,0}{\left(\theta\right)}=\frac{2h+2\left( R+h\right){\sin} \left(\left(\frac{\theta}{2}\right) \left(\frac{\pi}{180}\right)\right)}{2{\pi}R \left(\frac{\theta}{360}\right)\left(i\right)},
\end{equation}
where $\theta$ is the angular spacing between points $A$ and $B$ on the surface of the Earth or satellites $X$ and $Y$ in an OWSN in space, $h$ is the altitude of satellites in the OWSN, $R$ is the radius of the Earth, $i$ is the refractive index of the optical fiber in an OFTN, and 0 in $f_{crossover,0}$ indicates 0 intermediate hops, i.e., satellites $X$ and $Y$ are directly connected, as shown in Fig. 1. Note that points $A$ and $B$ could be optical fiber relays stations in an OFTN or satellite ground stations in an OWSN on the surface of the Earth. As shown in Fig. 1, satellites $X$ and $Y$ are exactly above points $A$ and $B$ in this scenario.

\begin{figure}[htbp]
	\centerline{\includegraphics[scale=0.47]{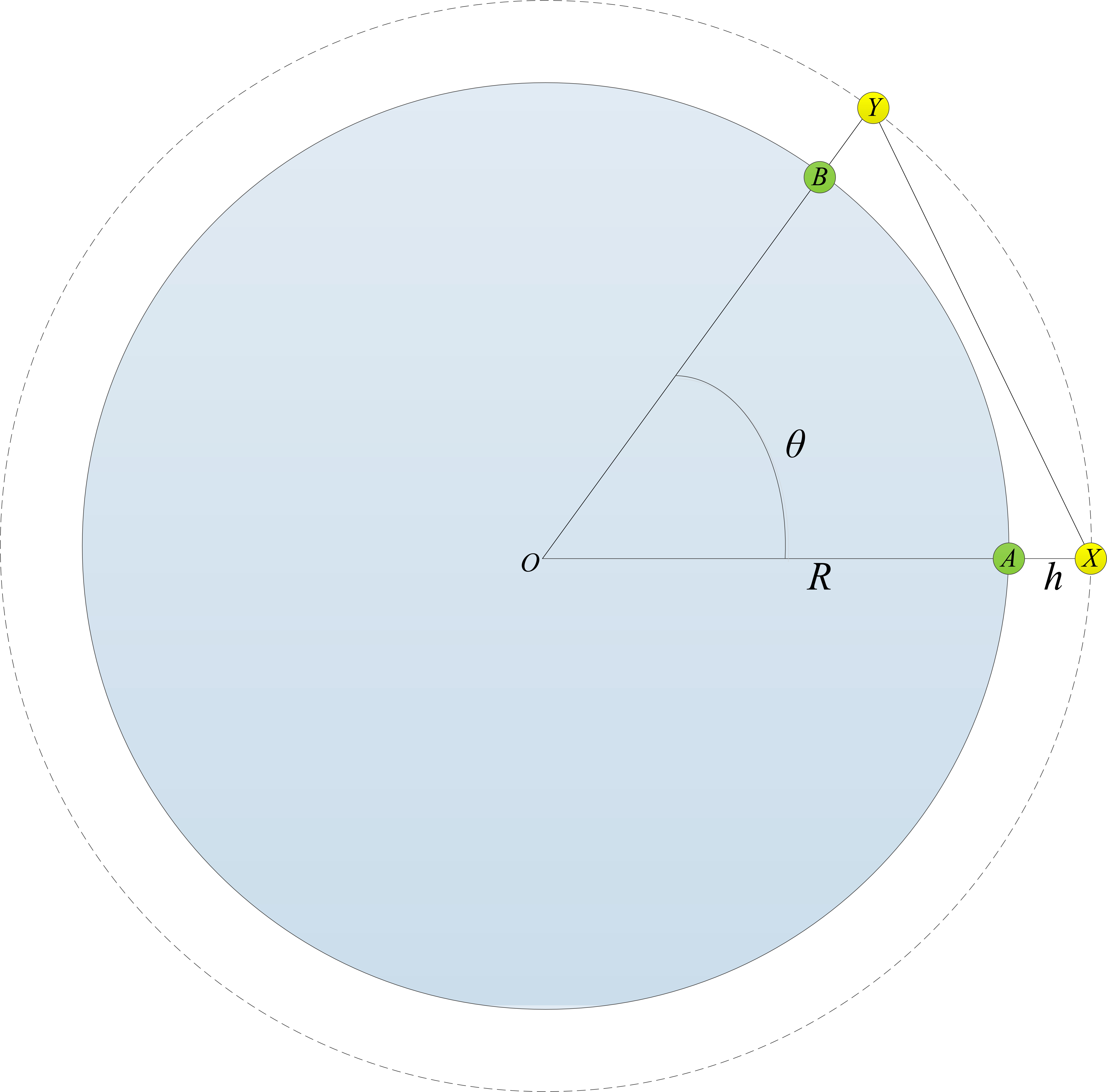}}
	\caption{Illustration of the first scenario in \cite{b14}, where the ingress and egress satellites $X$ and $Y$ in the OWSN are directly connected, $\theta$ is the angular spacing between points $A$ and $B$ on the surface of the Earth or satellites $X$ and $Y$ in the OWSN, $h$ is the altitude of satellites in the OWSN, $R$ is the radius of the Earth, and $O$ is the center of the Earth. The Earth is shown in blue and the dashed line represents the orbit of the satellites. The propagation distance between $X$ and $Y$ is equal to the length of chord $XY$.}
	\label{fig1}
\end{figure}

\par The end-to-end propagation distance between $A$ and $B$ over the OFTN (i.e., the length of the arc $AB$) in this scenario is

\begin{equation} \tag{2}
	\label{eq_2}
	d_\mathsf{{AB\_OFTN}}=2{\pi}R \left(\frac{\theta}{360}\right), 
\end{equation} 
and the end-to-end latency over the OFTN is
\begin{equation} \tag{3}
	\label{eq_3}
	t_\mathsf{{AB\_OFTN}}=2{\pi}R \left(\frac{\theta}{360}\right)\left(\frac{i}{c}\right), 
\end{equation}
where $c$ is the speed of light in vacuum. The end-to-end propagation distance between $A$ and $B$ over the OWSN (i.e., the altitude of ingress plus the length of chord $XY$ plus the altitude of egress) is
\begin{equation} \tag{4}
	\label{eq_4}
	d_\mathsf{{AB\_OWSN}}=2h+2\left( R+h\right){\sin} \left(\left(\frac{\theta}{2}\right) \left(\frac{\pi}{180}\right) \right),
\end{equation}
and the end-to-end latency over the OWSN is
\begin{equation} \tag{5}
	\label{eq_5}
	t_\mathsf{{AB\_OWSN}}=\frac{2h+2\left( R+h\right){\sin} \left(\left(\frac{\theta}{2}\right) \left(\frac{\pi}{180}\right)\right)}{c}.
\end{equation}

The crossover function in (1) is the ratio of the end-to-end latencies (or end-to-end propagation delays) over the OWSN and OFTN. The value of $\theta$ at which the value of the crossover function in (1) is equal to 1 is called crossover $\theta$ or $\theta_{crossover}$. The value of $\theta_{crossover}$ is substituted in (2) to determine the value of the crossover distance or $d_{crossover}$. Note that the propagation distance between satellites $X$ and $Y$, i.e., the length of chord $XY$ (or the length of the LISL between satellites $X$ and $Y$ when they are directly connected), is
\begin{equation} \tag{6}
	\label{eq_6}
	d_\mathsf{{XY}}=d_{LISL,0}=2\left( R+h\right){\sin} \left(\left(\frac{\theta}{2}\right) \left(\frac{\pi}{180}\right) \right).
\end{equation}

\par If there is one intermediate satellite, $C$, between satellites $X$ and $Y$, as shown in Fig. 2, then the length of the LISL between $X$ and $C$ or the length of the LISL between $C$ and $Y$, is
\begin{equation} \tag{7}
	\label{eq_7}
	d_\mathsf{{XC}}=d_\mathsf{{CY}}=d_{LISL,1}=2\left( R+h\right){\sin} \left(\left(\frac{\theta}{4}\right) \left(\frac{\pi}{180}\right) \right),
\end{equation}
when assuming that the length of chord $XC$ is the same as the length of chord $CY$. The crossover distance for this scenario can then be written as in (8).

\begin{figure}[htbp]
	\centerline{\includegraphics[scale=0.47]{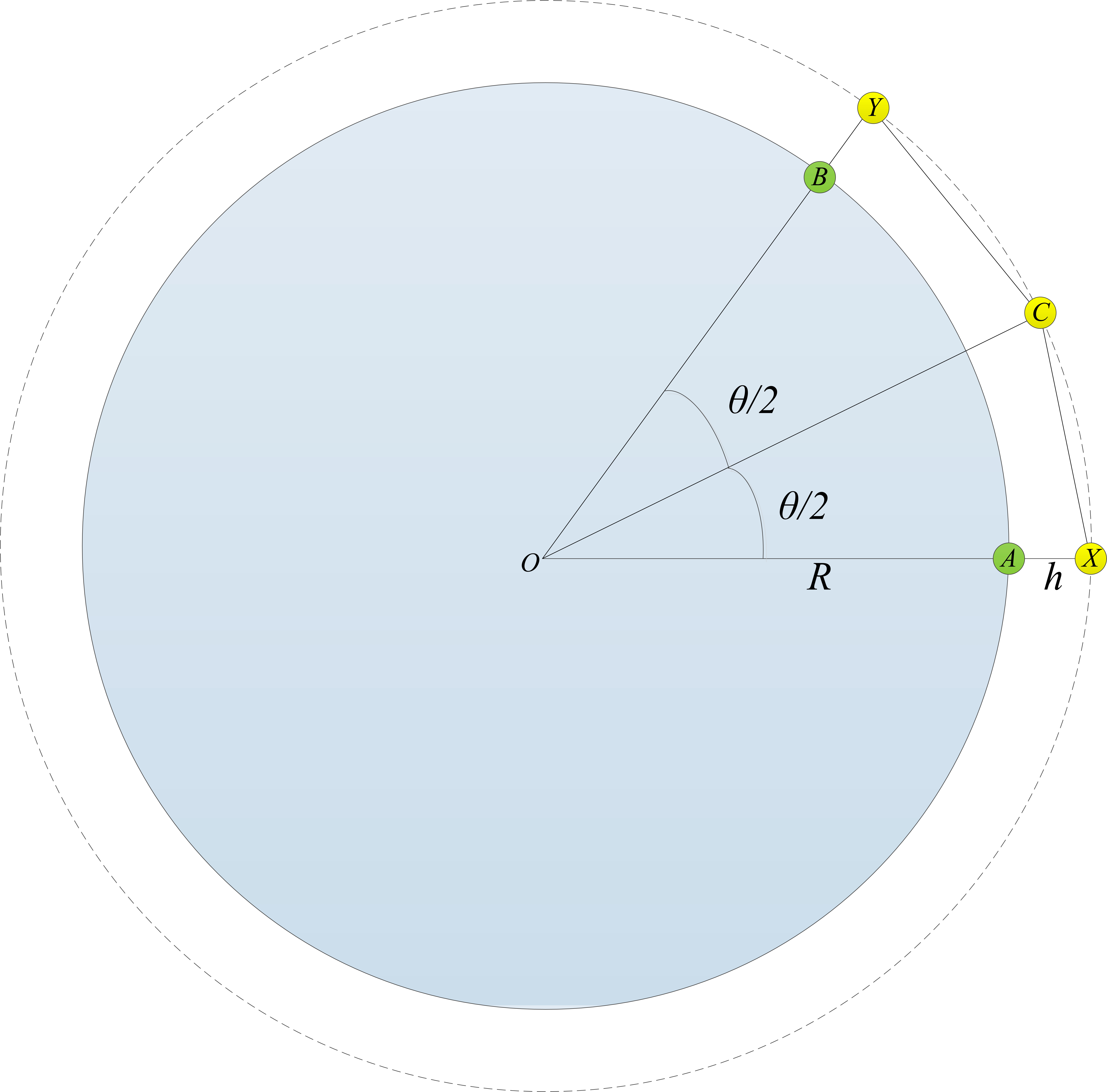}}
	\caption{Illustration of the scenario when satellites $X$ and $Y$ in the OWSN are connected via an intermediate satellite $C$. The propagation distance between $X$ and $Y$ is the length of chord $XC$ plus the length of chord $CY$.}
	\label{fig2}
\end{figure}

\par Similarly, the crossover distance for this scenario for the case of two intermediate satellites (see Fig. 3) between ingress and egress can be written as in (9) and the length of a LISL between a pair of satellites in this case is given by
\begin{equation} \tag{10}
	\label{eq_10}
	d_{LISL,2}=2\left( R+h\right){\sin} \left(\left(\frac{\theta}{6}\right) \left(\frac{\pi}{180}\right) \right).
\end{equation}
In general, the crossover function for this scenario in terms of $N$ intermediate satellites (or hops) between ingress and egress can be expressed as in (11) and the length of a LISL in terms of $N$ hops can be written as
\begin{equation} \tag{12}
	\label{eq_12}
	d_{LISL,N}=2\left( R+h\right){\sin} \left(\left(\frac{\theta}{2(N+1)}\right) \left(\frac{\pi}{180}\right) \right).
\end{equation}

\vspace{0.2cm}

\begin{figure}[htbp]
	\centerline{\includegraphics[scale=0.47]{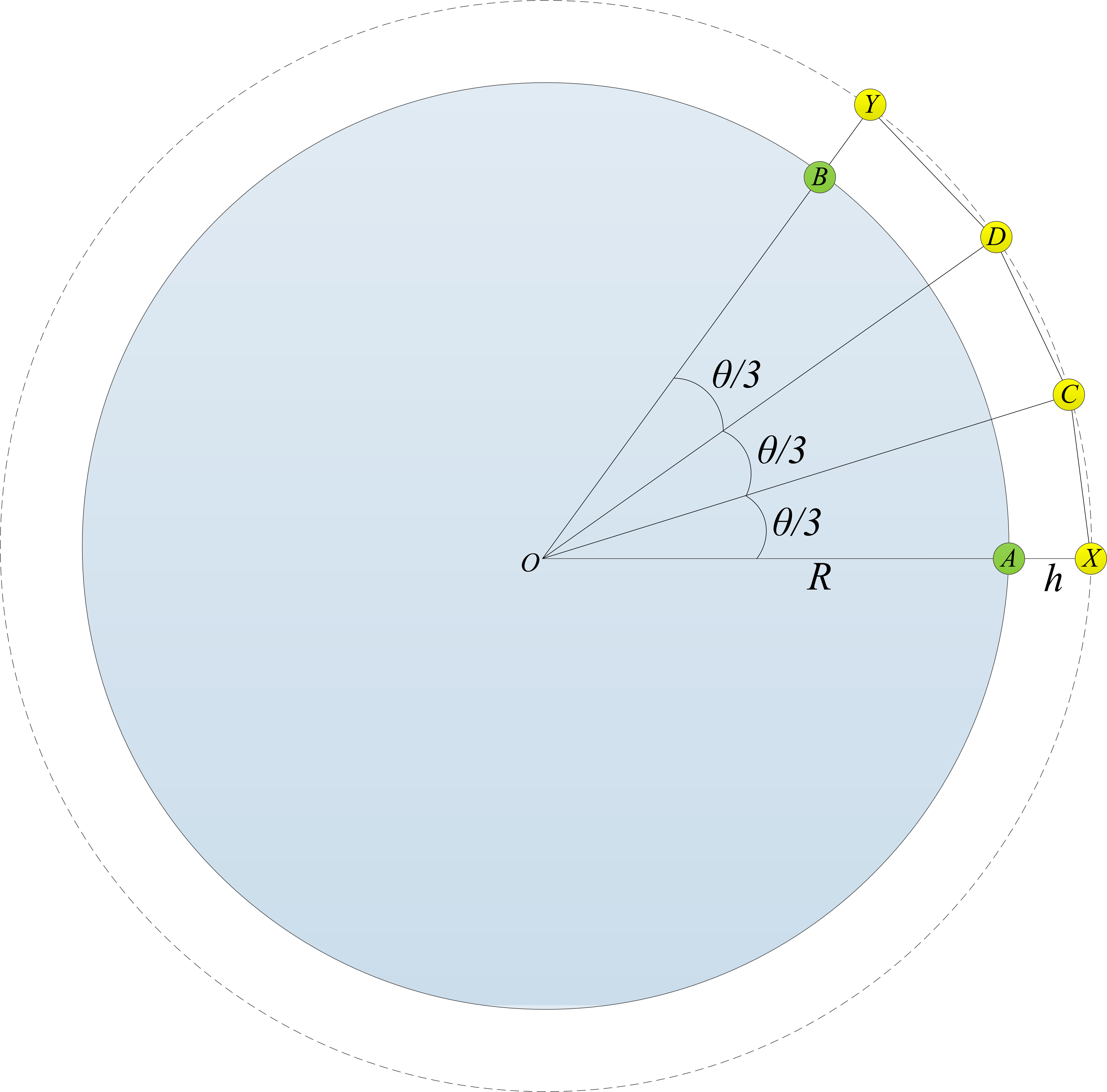}}
	\caption{Illustration of the scenario when satellites $X$ and $Y$ in the OWSN are connected via two intermediate satellites $C$ and $D$. The propagation distance between $X$ and $Y$ is sum of the lengths of chords $XC$, $CD$, and $DY$.}
	\label{fig3}
\end{figure}

\begin{table*}[htbp]
\vspace{0.2cm}
	\fontsize{10}{12}\selectfont
\begin{equation} \tag{8}
	\label{eq_8}
	f_{crossover,1}{\left(\theta\right)}=\frac{2h+2\left( R+h\right){\sin} \left(\left(\frac{\theta}{4}\right) \left(\frac{\pi}{180}\right)\right)+2\left( R+h\right){\sin} \left(\left(\frac{\theta}{4}\right) \left(\frac{\pi}{180}\right)\right)}{2{\pi}R \left(\frac{\theta}{360}\right)\left(i\right)}
\end{equation}
\vspace{1cm}
\begin{equation} \tag{9}
	\label{eq_9}
	f_{crossover,2}{\left(\theta\right)}=\frac{2h+2\left( R+h\right){\sin} \left(\left(\frac{\theta}{6}\right) \left(\frac{\pi}{180}\right)\right)+2\left( R+h\right){\sin} \left(\left(\frac{\theta}{6}\right) \left(\frac{\pi}{180}\right)\right)+2\left( R+h\right){\sin} \left(\left(\frac{\theta}{6}\right) \left(\frac{\pi}{180}\right)\right)}{2{\pi}R \left(\frac{\theta}{360}\right)\left(i\right)}
\end{equation}
\vspace{1cm}
\begin{equation} \tag{11}
	\label{eq_11}
	f_{crossover,N}{\left(\theta\right)}=\frac{2h+(N+1)\left(2\left( R+h\right){\sin} \left(\left(\frac{\theta}{2(N+1)}\right) \left(\frac{\pi}{180}\right)\right)\right)}{2{\pi}R \left(\frac{\theta}{360}\right)\left(i\right)}
\end{equation}
\hrulefill
\end{table*}

\section{Numerical Results}

\par To evaluate the effect of different number of hops on the crossover distance and length of a LISL in this scenario, we consider three different OWSNs with different satellite altitudes and three different OFTNs with different optical fiber refractive indexes. The altitude of satellites (or $h$) in the three OWSNs is considered as 300 km, 550 km, and 1,100 km, respectively, whereas the refractive index of the optical fiber (or $i$) in the three OFTNs is taken as 1.4675, 1.3, and 1.1, respectively. Note that 550 km is the altitude of satellites in the satellite constellation for Phase I of Starlink [17]. Other upcoming VLEO/LEO satellite constellations may have different altitudes and so we consider OWSNs with different satellite altitudes as well. It should also be noted that the refractive index of long-distance submarine optical fiber cables is 1.4675 [18]. Technological developments in future may lead to submarine optical fiber cables with lower refractive indexes and, thereby, we also consider OFTNs with lower refractive indexes. The radius of the Earth is taken as 6,378 km.  

\par Table 1 shows crossover $\theta$, crossover distance, and length of a LISL at different values of $h$, $i$, and $N$, where $N$ is the number of hops or intermediate satellites between the ingress and egress satellites. In this table, $\theta_{c}$ and $d_{c}$ denote $\theta_{crossover}$ and $d_{crossover}$, respectively, whereas $d_{L}$ represents $d_{LISL}$ or length of a LISL. We also plot $d_{crossover}$ and $d_{LISL}$ versus $N$ at different $h$ and $i$ in Figs. 4 and 5, respectively, so that trends are easily apparent. In these figures, the solid lines with circle markers, the dashed lines with square markers, and the dashed-dotted lines with diamond markers represent $d_{crossover}$ and $d_{LISL}$ at $i$ = 1.4675, 1.3, and 1.1, respectively, while the blue, red, and black colored lines in these figures indicate $d_{crossover}$ and $d_{LISL}$ at $h$ = 300 km, 550 km, and 1,100 km, respectively.

\begin{table*}
	\fontsize{8}{13}\selectfont
	\centering
	\renewcommand\thetable{1}\caption{$\theta_{c}$, $d_{c}$, and $d_{L}$ for different values of $h$, $i$, and $N$ are shown. Here, $\theta_{c}$, $d_{c}$, and $d_{L}$ represent $\theta_{crossover}$ (i.e., crossover $\theta$), $d_{crossover}$ (i.e., crossover distance), and $d_{LISL}$ (i.e., length of a LISL), respectively.}
	\begin{tabular}{|c|c|c|c|c|c|c|c|c|c|c|} 
		\hline
		\multirow{2}{*}{\begin{tabular}[c]{@{}c@{}}\textbf{\textit{h}} \\\textbf{(km)\textit{}}\end{tabular}} & \multirow{2}{*}{\textbf{\textit{i}}} & \multicolumn{3}{c|}{\textbf{\textit{N = }0\textit{ }}}                                                                                                                                                                                                                                                       & \multicolumn{3}{c|}{\textbf{\textit{N = }1\textit{}}}                                                                                                                                                                                                                                                                          & \multicolumn{3}{c|}{\textbf{\textit{N = }2\textit{}}}                                                                                                                                                                                                                                                                           \\ 
		\cline{3-11}
		&                                      & \begin{tabular}[c]{@{}c@{}}\boldmath{$\theta_{c,0}$} \\\textbf{(degrees)}\end{tabular} & \begin{tabular}[c]{@{}c@{}}\boldmath{$d_{c,0}$} \\\textbf{(km)}\end{tabular} & \begin{tabular}[c]{@{}c@{}}\boldmath{$d_{L,0}$} \\\textbf{(km)\textit{}}\end{tabular} & \begin{tabular}[c]{@{}c@{}}\boldmath{$\theta_{c,1}$} \\\textbf{(degrees)\textit{}}\end{tabular} & \begin{tabular}[c]{@{}c@{}}\boldmath{$d_{c,1}$} \\\textbf{(km)\textit{}}\end{tabular} & \begin{tabular}[c]{@{}c@{}}\boldmath{$d_{L,1}$} \\\textbf{(km)\textit{}}\end{tabular} & \begin{tabular}[c]{@{}c@{}}\boldmath{$\theta_{c,2}$} \\\textbf{(degrees)\textit{}}\end{tabular} & \begin{tabular}[c]{@{}c@{}}\boldmath{$d_{c,2}$} \\\textbf{(km)\textit{}}\end{tabular} & \begin{tabular}[c]{@{}c@{}}\boldmath{$d_{L,2}$} \\\textbf{(km)\textit{}}\end{tabular}  \\ 
		\hline
		\multirow{3}{*}{300}                                                                                  & 1.4675                               & 12.7523                                                                                              & 1,420                                                                                           & 1,483                                                                                               & 12.8023                                                                                                       & 1,425                                                                                                    & 746                                                                                                 & 12.8123                                                                                                       & 1,426                                                                                                    & 498                                                                                                  \\ 
		\cline{2-11}
		& 1.3                                  & 20.8311                                                                                              & 2,319                                                                                           & 2,415                                                                                               & 21.1836                                                                                                       & 2,358                                                                                                    & 1,233                                                                                               & 21.2521                                                                                                       & 2,366                                                                                                    & 825                                                                                                  \\ 
		\cline{2-11}
		& 1.1                                  & 56.6665                                                                                              & 6,308                                                                                           & 6,339                                                                                               & 75.2285                                                                                                       & 8,374                                                                                                    & 4,306                                                                                               & 84.8275                                                                                                       & 9,443                                                                                                    & 3,262                                                                                                \\ 
		\hline
		\multirow{3}{*}{550}                                                                                  & 1.4675                               & 25.3295                                                                                              & 2,820                                                                                           & 3,038                                                                                               & 25.7646                                                                                                       & 2,868                                                                                                    & 1,554                                                                                               & 25.8496                                                                                                       & 2,878                                                                                                    & 1,041                                                                                                \\ 
		\cline{2-11}
		& 1.3                                  & 41.6112                                                                                              & 4,632                                                                                           & 4,922                                                                                               & 44.7825                                                                                                       & 4,985                                                                                                    & 2,690                                                                                               & 45.5536                                                                                                       & 5,071                                                                                                    & 1,831                                                                                                \\ 
		\cline{2-11}
		& 1.1                                  & 86.5597                                                                                              & 9,636                                                                                           & 9,499                                                                                               & 133.4093                                                                                                      & 14,851                                                                                                   & 7,618                                                                                               & 170.7591                                                                                                      & 19,008                                                                                                   & 6,603                                                                                                \\ 
		\hline
		\multirow{3}{*}{1,100}                                                                                & 1.4675                               & 57.5088                                                                                              & 6,402                                                                                           & 7,195                                                                                               & 63.7364                                                                                                       & 7,095                                                                                                    & 4,106                                                                                               & 65.4245                                                                                                       & 7,283                                                                                                    & 2,829                                                                                                \\ 
		\cline{2-11}
		& 1.3                                  & 85.0740                                                                                              & 9,470                                                                                           & 10,111                                                                                              & 113.1893                                                                                                      & 12,600                                                                                                   & 7,090                                                                                               & 127.9898                                                                                                      & 14,247                                                                                                   & 5,441                                                                                                \\ 
		\cline{2-11}
		& 1.1                                  & 127.5217                                                                                             & 14,195                                                                                          & 13,415                                                                                              & 214.9175                                                                                                      & 23,924                                                                                                   & 12,058                                                                                              & 295.6886                                                                                                      & 32,915                                                                                                   & 11,335                                                                                               \\
		\hline
	\end{tabular}
\end{table*}    

\par As can be seen from Table 1 and Fig. 4, the crossover distance increases with an increase in $N$ at different $h$ and $i$. For example, $d_{crossover}$ is 1,420 km, 1,425 km, and 1,426 km at $h$ = 300 km and $i$ = 1.4675; it is 4,632 km, 4,985 km, and 5,071 km at $h$ = 550 km and $i$ = 1.3; and it is 14,195 km, 23,924 km, and 32,915 km at $h$ = 1,100 km and $i$ = 1.1 for $N$ = 0, 1, and 2, respectively. The increase in the crossover distance with an increase in $N$ is more significant at higher $h$ and lower $i$. As observed from Table 1 and Fig. 5, the length of a LISL decreases with an increase in $N$ at different $h$ and $i$. For example, $d_{LISL}$ is 1,483 km, 746 km, and 498 km at $h$ = 300 km and $i$ = 1.4675; it is 4,922 km, 2,690 km, and 1,831 km at $h$ = 550 km and $i$ = 1.3; and it is 13,415 km, 12,058 km, and 11,335 km at $h$ = 1,100 km and $i$ = 1.1 for $N$ = 0, 1, and 2, respectively.

\par Note that the crossover distance depends upon the propagation distance between ingress and egress satellites in an OWSN. This propagation distance (shown as $d_{XY}$ in Fig. 1 for $N$ = 0, $d_{XC}$ + $d_{CY}$ in Fig. 2 for $N$ = 1, and $d_{XC}$ + $d_{CD}$ + $d_{DY}$ in Fig. 3 for $N$ = 2) is equal to the length of a LISL times the number of LISLs between ingress and egress. For $N$ = 0, 1, and 2, the number of LISLs is 1, 2, and 3, respectively. The propagation distance increases with an increase in $N$ and is more at higher $h$ and lower $i$. For example, at $h$ = 300 km and $i$ = 1.4675, the propagation distance between ingress and egress is 1,483 km at $N$ = 0, it is 1,492 km (i.e., 746 km times 2) at $N$ = 1, and it is 1,494 km (i.e., 498 km times 3) at $N$ = 2. The increase in the propagation distance between $X$ and $Y$ with an increase in $N$ at a certain $h$ and $i$ results in an increase in the crossover distance with an increase in $N$ at that $h$ and $i$. An increase in $N$, i.e., an increase in the number of intermediate satellites, means an increase in the number of LISLs. More LISLs mean more indirect LISLs between $X$ and $Y$, which translate into a higher propagation distance between $X$ and $Y$ that in turn results in a higher crossover distance. The increase in the crossover distance with an increase in $N$ is higher at higher values of $h$ and lower values of $i$ as the increase in the propagation distance between $X$ and $Y$ with an increase in $N$ is higher at these $h$ and $i$. For example, the propagation distance between $X$ and $Y$ is 1,483 km, 1,492 km, and 1,494 km at $h$ = 300 km and $i$ = 1.4675; it is 4,922 km, 5,380 km, and 5,493 km at $h$ = 550 km and $i$ = 1.3; and it is 13,415 km, 24,116 km, and 34,005 km at $h$ = 1,100 km and $i$ = 1.1 for $N$ = 0, 1, and 2, respectively. 

\par It is interesting to note that the crossover distance and length of a LISL have an inverse relationship, as can be seen from Figs. 4 and 5. As $N$ increases, we can see from these figures that $d_{crossover}$ increases, while $d_{LISL}$ decreases. For example, at $h$ = 550 km and $i$ = 1.4675, $d_{crossover}$ is 2,820 km, 2,868 km, and 2,878 km, while $d_{LISL}$ is 3,038 km, 1,554 km, and 1,041 km for $N$ = 0, 1, and 2, respectively. For a certain value of $h$ and $i$, as $N$ increases, the number of intermediate satellites increases and, thereby, the number of LISLs increases. This results in shorter LISLs and the length of a LISL decreases, which means more indirect LISLs between $X$ and $Y$. This leads to an increase in the propagation distance between ingress and egress, which translates into an increase in the crossover distance.

\begin{figure}[htbp]
	\centerline{\includegraphics[scale=0.098]{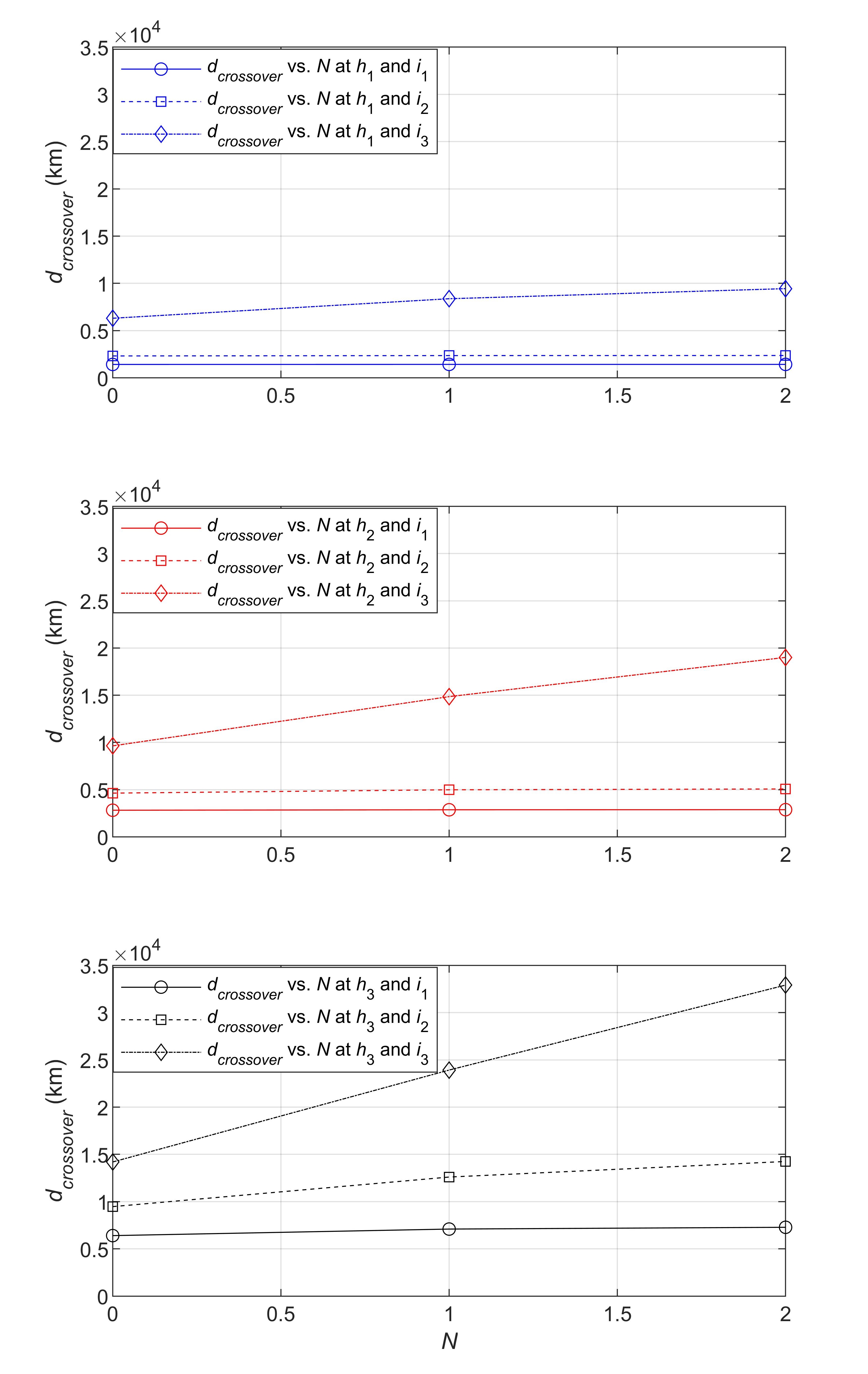}}
	\caption{Crossover distance (km) vs. number of hops for different values of $h$ and $i$ is shown. In these plots, $h_1$, $h_2$, and $h_3$ indicate satellite altitudes of 300 km, 550 km, and 1,100 km, respectively, in the three different OWSNs, whereas $i_1$, $i_2$, and $i_3$ represent optical fiber refractive indexes of 1.4675, 1.3, and 1.1, respectively, in the three different OFTNs.}
	\label{fig2}
\end{figure}

\begin{figure}[htbp]
	\centerline{\includegraphics[scale=0.098]{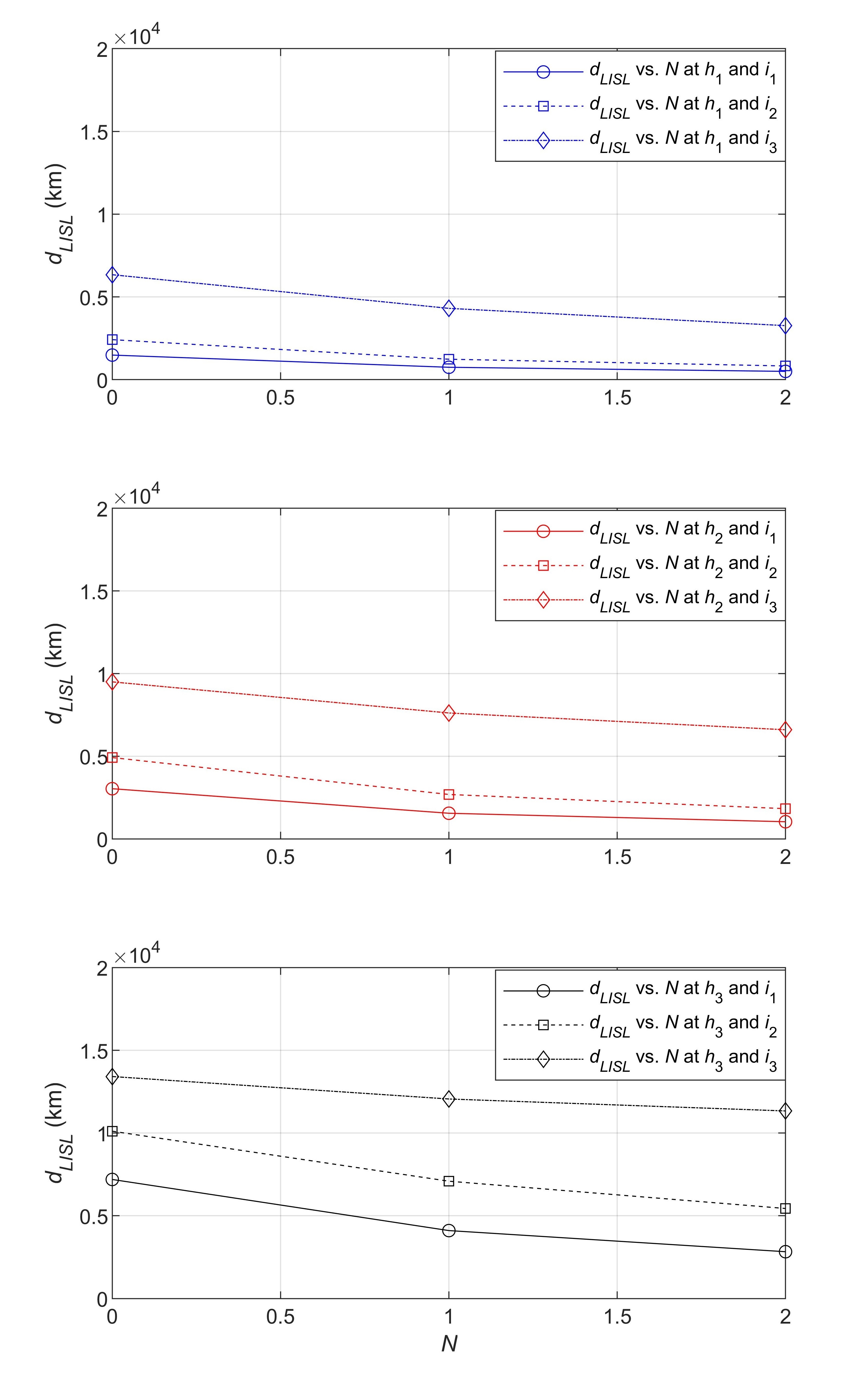}}
	\caption{Length of a LISL (km) vs. number of hops for different values of $h$ and $i$ is shown. In these plots, $h_1$, $h_2$, and $h_3$ indicate satellite altitudes of 300 km, 550 km, and 1,100 km, respectively, in the three different OWSNs, whereas $i_1$, $i_2$, and $i_3$ represent optical fiber refractive indexes of 1.4675, 1.3, and 1.1, respectively, in the three different OFTNs.}
	\label{fig2}
\end{figure}

\section{Conclusions}

\par In this work, we extend our previously proposed crossover function for a scenario to include intermediate satellites between the ingress and egress satellites in an OWSN to find a more realistic value of the crossover distance for switching from an OFTN to an OWSN for lower latency data communications between two points, such as optical fiber relay stations or satellite ground stations, on Earth. We investigate the effect of the number of hops (or intermediate satellites) between ingress and egress satellites on the crossover distance and length of a LISL in three different OWSNs with different satellite altitudes and three different OFTNs with different optical fiber refractive indexes. 

\par Numerical results indicate that the crossover distance increases, whereas the length of a LISL decreases with an increase in the number of hops. For a certain satellite altitude in an OWSN and optical fiber refractive index in an OFTN, the number of LISLs increases as the number of hops increases. Consequently, the length of a LISL decreases, and this creates shorter, more, and indirect LISLs leading to an increase in the propagation distance between ingress and egress in the OWSN, which in turn results in an increase in the crossover distance. Furthermore, a higher increase in the crossover distance with an increase in the number of hops is observed with OWSNs at higher satellite altitudes and OFTNs at lower refractive indexes since the increase in propagation distance between ingress and egress is higher for these OWSNs and OFTNs. 

\par The crossover functions for four different scenarios have been proposed in \cite{b14}, while in this work, we extend and evaluate the crossover function for the first scenario in \cite{b14}. In the future, we plan to extend the crossover functions for the other scenarios as well to incorporate intermediate satellites between ingress and egress in these scenarios for obtaining a more realistic crossover distance in these scenarios.

\section*{Acknowledgment}

{\par This work was supported by the High Throughput and Secure Networks Challenge Program at the National Research Council of Canada. The authors would like to thank AGI for the STK platform.}

\end{document}